\documentclass{article}
\usepackage{spconf,amsmath,graphicx}
\usepackage{url}
\usepackage{multirow}
\usepackage{amsmath}
\usepackage{amssymb}

\title{ToneUnit: A Speech Discretization Approach for Tonal Language Speech Synthesis}
\name{Dehua Tao$^{1,*}$\thanks{* This work was done during an internship at Huawei Noah’s Ark Lab.}, Daxin Tan$^2$, Yu Ting Yeung$^2$, Xiao Chen$^2$, Tan Lee$^1$}
\address{
  $^1$ Department of Electronic Engineering, The Chinese University of Hong Kong \\ 
   $^2$ Huawei Noah’s Ark Lab \\
   dhtao@link.cuhk.edu.hk, tanlee@ee.cuhk.edu.hk,\\ \{tandaxin1, yeung.yu.ting, chen.xiao2\}@huawei.com
}

\begin{document}
% \ninept
%
\maketitle
\begin{abstract}
% The abstract should appear at the top of the left-hand column of text, about
% 0.5 inch (12 mm) below the title area and no more than 3.125 inches (80 mm) in
% length.  Leave a 0.5 inch (12 mm) space between the end of the abstract and the
% beginning of the main text.  The abstract should contain about 100 to 150
% words, and should be identical to the abstract text submitted electronically
% along with the paper cover sheet.  All manuscripts must be in English, printed
% in black ink.
Representing speech as discretized units has numerous benefits in supporting downstream spoken language processing tasks. However, the approach has been less explored in speech synthesis of tonal languages like Mandarin Chinese. Our preliminary experiments on Chinese speech synthesis reveal the issue of ``tone shift", where a synthesized speech utterance contains correct base syllables but incorrect tones. To address the issue, we propose the ToneUnit framework, which leverages annotated data with tone labels as CTC supervision to learn tone-aware discrete speech units for Mandarin Chinese speech. Our findings indicate that the discrete units acquired through the TonUnit resolve the ``tone shift" issue in synthesized Chinese speech and yield favorable results in English synthesis. Moreover,  the experimental results suggest that finite scalar quantization enhances the effectiveness of ToneUnit. Notably, ToneUnit can work effectively even with minimal annotated data.\footnote{The synthesized speech samples are provided at \url{https://toneunit1225.github.io/}}
\end{abstract}
\section{Introduction}
\label{sec:intro}

%Discrete speech representation learning has become a recent research interest. The information of a speech segment in a short time interval is represented by a single token (speech unit) instead of a high-dimensional continuous vector. A benefit of applying speech units is a significant reduction of data size for storage and transmission \cite{chang2023exploration}. The speech unit sequence can be further shortened by techniques like de-duplication \cite{lee2021direct} and sub-word modeling \cite{hayashi2020discretalk}, to accelerate the training and inference processes. Moreover, the analogy between discrete speech units and text tokens opens up the potential to apply various kinds of Natural Language Processing (NLP) techniques to speech processing tasks \cite{baevski2019effectiveness,hsu2021hubert,chung2021w2v,lakhotia2021generative,borsos2023audiolm}.

Discrete speech representation learning has become a recent research interest. The information of a speech segment in a short time interval is represented by a single token, namely a speech unit, instead of a high-dimensional continuous vector. Discrete units facilitate the storage and transmission of speech data. The analogy between discrete speech units and text tokens opens up the potential to apply Natural Language Processing (NLP) techniques to speech processing tasks \cite{baevski2019effectiveness,hsu2021hubert,chung2021w2v,lakhotia2021generative,borsos2023audiolm}. To obtain discrete units, self-supervised learning (SSL) speech foundation model and discretization methods are adopted. SSL models leverage large amounts of unlabelled speech data to derive continuous representations that contain rich information in speech signals \cite{mohamed2022self}. Subsequently, discretization methods convert the learned continuous representations into discrete units.

% https://ieeexplore.ieee.org/stamp/stamp.jsp?arnumber=6707740&tag=1
% https://ieeexplore.ieee.org/stamp/stamp.jsp?arnumber=4284659
% Pitch Ability As an Aptitude for Tone Learning (see Lexical Tone)
While considerable progress has been made in developing unit-based speech models primarily for non-tonal languages like English, the exploration of tonal languages such as Mandarin Chinese is still limited. Linguistically, tone refers to the use of pitch in determining the meaning of a word \cite{yip2002tone}. Tonal languages use tones to differentiate phones and words. This is crucial to lexical and grammatical differentiation \cite{bao1999structure}. Four lexical tones are used in Mandarin Chinese: High (Tone 1), Rising (Tone 2), Low / Dipping (Tone 3), and Falling (Tone 4). The tone of a syllable is carried out by the vowel nucleus. Our preliminary experiments reveal the issue of ``tone shift" in synthesized Mandarin Chinese speech with discrete speech units. ``Tone shift" is said to occur when a synthesized speech utterance contains correct base syllables but incorrect tones, causing misunderstanding of the speech content and degradation of speech intelligibility.

%While considerable progress has been made in developing speech unit for non-tonal languages like English, the exploration of tonal languages such as Mandarin is still limited. There exists a large gap between tonal language and non-tonal language in terms of phonetic and morphological structures. In tonal languages, tones are used to differentiate phones and words, thus play a vital role in the determination of meaning\cite{bao1999structure}. In Mandarin, four tones are used: High (Tone 1), Rising (Tone 2), Low/Dipping (Tone 3), and Falling (Tone 4). Our preliminary experiments reveal the ``tone shift" problem in the Mandarin speech synthesized from discrete speech units. ``Tone shift" is said to occur when a synthesized speech utterance contains correct base syllables but incorrect tones. This would cause misunderstanding of speech content and degradation of speech intelligibility. 

\begin{figure}[!t]

\begin{minipage}[b]{.45\linewidth}
  \centering
  \centerline{\includegraphics[width=4.0cm]{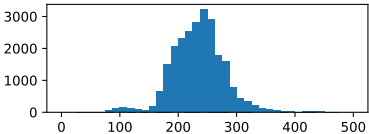}}
  \centerline{Tone 1 of /i/}\medskip
\end{minipage}
\hfill
\begin{minipage}[b]{.45\linewidth}
  \centering
  \centerline{\includegraphics[width=4.0cm]{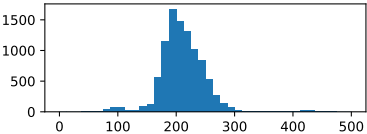}}
  \centerline{Tone 2 of /i/}\medskip
\end{minipage}

\begin{minipage}[b]{.45\linewidth}
  \centering
  \centerline{\includegraphics[width=4.0cm]{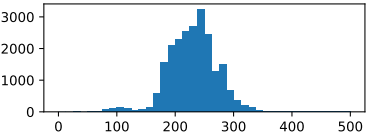}}
  \centerline{Tone 4 of /i/}\medskip
\end{minipage}
\hfill
\begin{minipage}[b]{.45\linewidth}
  \centering
  \centerline{\includegraphics[width=4.0cm]{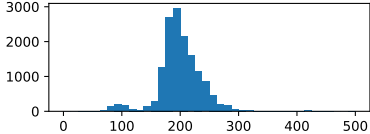}}
  \centerline{Tone 3 of /i/}\medskip
\end{minipage}

\caption{Pitch distribution for four tones of phoneme /i/ in Mandarin Chinese speech. The x-axis represents fundamental frequency, and the y-axis is the number of occurrences.}
\label{fig:tonalph}
\end{figure}

We hypothesize that the ``tone shift'' issue stems from the generation process of discrete speech units. An SSL model with only speech self-supervision and an unsupervised clustering method like k-means are typically adopted to obtain speech units. Since tone is highly related to pitch, we analyze the relationship between the tones of Mandarin Chinese and the distribution of pitch values in natural speech. It is found that a similarity exists between Tone 1 and Tone 4, as well as between Tone 2 and Tone 3. Figure \ref{fig:tonalph} illustrates such similarity with the vowel /i/ as an example. This observation suggests that tonal information can not be adequately captured from speech signals. When discrete speech units are derived solely from acoustic signals without the guidance of linguistic information, they may not be able to reproduce the desired tones accurately in synthesized speech. This may explain that when applying unsupervised clustering to latent speech representations derived by SSL, the resulting units may not represent tonal variations of the same phoneme, though they generally suffice to differentiate the phonemes.

The present study introduces a speech discretization framework to generate tone-aware speech units that can effectively capture lexical tone information in Mandarin Chinese. This framework, called \textit{ToneUnit}, comprises an SSL-based speech encoder, a quantization module, and a tonal phone decoder. The speech encoder is pre-trained on a large amount of unlabeled speech data. It is fine-tuned jointly with the quantization module and the tonal phone decoder on a small amount of annotated data. Specifically, phone sequences with tone markers serve as the target labels for fine-tuning the entire framework with Connectionist Temporal Classification (CTC) loss, thus compelling the quantizer to generate speech units that model tone variations. Two quantization methods are investigated, including Gumbel-Softmax-based vector quantization (VQ) \cite{van2017neural,baevski2019vq,baevski2020wav2vec} and finite scalar quantization (FSQ) \cite{mentzer2023finite}. Experimental results show that the ToneUnit framework can effectively address the ``tone shift" issue in Mandarin Chinese speech synthesis, meanwhile demonstrating competitive performance in English speech synthesis. Moreover, FSQ exhibits superior performance to VQ with a more straightforward architecture and improved codebook utilization. To the best of our knowledge, this is the first attempt at applying FSQ to speech discretization.

This paper is organized as follows. In the next Section, we review related works on speech discretization. We describe the proposed ToneUnit framework in Section \ref{sec:method} and our experimental setup in Section \ref{sec:setup}. We discuss the experimental results in Section \ref{rec:result}. Finally, we conclude our work in Section \ref{sec:Conclusion}.

\section{Related Works}
% Boosting Self-Supervised Embeddings for Speech Enhancement (see SSL models)
\subsection{Self-supervised learning models}

By leveraging large quantities of unlabeled speech data, SSL models including wav2vec 2.0 \cite{baevski2020wav2vec}, HuBERT \cite{hsu2021hubert}, WavLM \cite{chen2022wavlm} and SPIRAL \cite{huang2022spiral}, have achieved remarkable performance in various speech-related downstream tasks \cite{yang2021superb}. SSL models are usually trained with contrastive loss (e.g., wav2vec 2.0, SPIRAL) or masked language model loss (e.g. HuBERT, WavLM). To further improve model robustness, SPIRAL relies on a teacher-student architecture while WavLM proposes a masked speech denoising and prediction framework during training. The latent representations derived from SSL models contain rich information of speech signals, such as content, speaker identity, style, and emotion. For example, WavLM is designed for full stack speech processing tasks. These representations can serve as alternatives to traditional speech features like Mel Frequency Cepstral Coefficients (MFCC) and log Mel filter banks (FBANK) for input into downstream models.

\subsection{Speech quantization} 

% The methods for speech discretization in the previous studies include vector quantization (VQ) and clustering. The VQ-based methods typically learn a specialized vector quantization module in the training stage, such as VQ-VAE [1], Wav2Vec models [2,3] with VQ module, and neural codec models [4,5] with residual VQ components. Alternatively, clustering algorithms are applied to the latent representation from a pre-trained speech encoder, where the cluster indices are used as discrete representations of speech signals. This approach can be seen in HuBERT-like models [6, 7], which often employ K-means clustering.

% % [1] Neural discrete representation learning
% % [2] vq-wav2vec:self-supervised learning of discrete speech representations
% % [3] wav2vec2.0: a framework for self-supervised learning of speech representations
% % [4] soundstream: a end-to-end neural audio codec
% % [5] high fidelity neural audio compression
% % [6] HuBERT: self-supervised speech representation learning by masked prediction of hidden units
% % [7] wavlm: large-scale of self-supervised pre-training for full stack speech processing

In previous studies, discrete speech units are typically obtained by employing VQ modules \cite{van2017neural} with SSL models \cite{baevski2019vq,baevski2020wav2vec}, or through k-means clustering applied to hidden embeddings of SSL models \cite{hsu2021hubert,chen2022wavlm}. The VQ codebook of a predetermined size is trained to project input vectors into the nearest codebook entry.  Optimizing the VQ codebook is challenging due to the codebook collapse problem \cite{dieleman2018challenge,baevski2019vq,dhariwal2020jukebox,takida2022sq}, which leads to poor reconstruction with only a few activated codewords. Finite scalar quantization (FSQ) \cite{mentzer2023finite} is proposed to solve the codebook collapse problem without relying on any auxiliary losses.

%In previous studies, discrete speech units are typically obtained by employing VQ modules \cite{van2017neural} with SSL models \cite{baevski2019vq,baevski2020wav2vec}, or through k-means clustering applied to hidden embeddings from these models \cite{hsu2021hubert,chen2022wavlm}. The codebook of VQ with a predetermined size is trained to project input vectors onto the nearest codebook entry. However, such codebook is challenging to optimize and suffers from the collapse problem \cite{dieleman2018challenge,baevski2019vq,dhariwal2020jukebox,takida2022sq}: only a few codewords are used, leading to poor reconstructions. Recently, FSQ \cite{mentzer2023finite} is proposed to solve the problem of codebook collapse without relying on any auxiliary losses.

\subsection{Speech synthesis from discrete units}

Discrete speech units have been extensively studied in speech generation tasks \cite{van2017neural,eloff2019unsupervised,lakhotia2021generative,lee2021textless,lee2021direct,polyak2021speech}. HiFi-GAN \cite{kong2020hifi} with a duration predictor \cite{ren2020fastspeech} can be used as a unit-based vocoder to decode speech signal from discrete speech representations \cite{lee2021textless,lee2021direct}. HiFi-GAN is a generative adversarial network (GAN) model that focuses on modeling periodic patterns in speech audio to enhance quality. %VITS \cite{kim2021conditional}, a text-to-speech end-to-end model, incorporates a stochastic duration predictor to generate speech with variable rhythms from the input text. %In this study, we employ VITS as the discrete unit-based speech synthesizer.
VITS \cite{kim2021conditional} is a parallel text-to-speech system designed to perform both learning and
synthesis in an end-to-end manner. VITS features a stochastic duration predictor to capture varied rhythms of speech that are unable to be represented by text.

\section{ToneUnit framework}
\label{sec:method}
%The overall structure of ToneUnit is depicted in Figure~\ref{fig:sys_struct}. The framework consists of three components: a pre-trained speech encoder, a quantizer module, and a CTC decoder. The speech encoder converts the input speech signals into continuous representations, which are fed into the quantizer. The quantized embeddings, i.e., codebook vectors, are passed to the decoder. The parameters of the three components are jointly updated during the fine-turning phase. Finally, a speech synthesizer is separately trained to convert discrete units generated by the quantizer into speech signals.

\begin{figure}[t]
  \centering
  \includegraphics[width=0.9\linewidth]{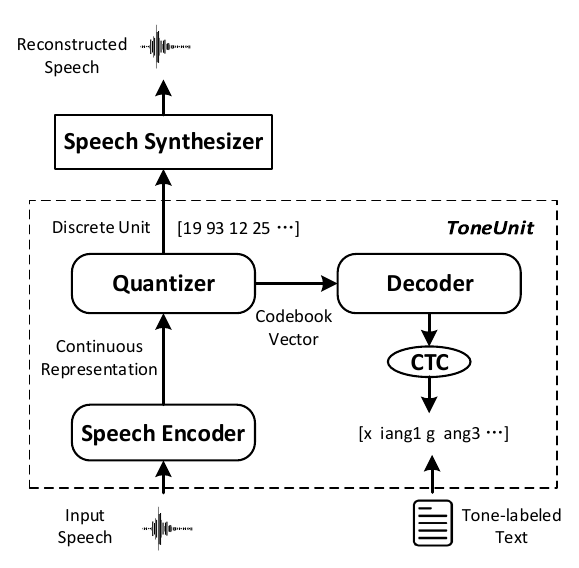}
  \caption{Components of the ToneUnit. The speech synthesizer is trained independently on discrete speech units generated by ToneUnit.}
  \label{fig:sys_struct}
\end{figure}

The overall structure of ToneUnit is depicted in Figure~\ref{fig:sys_struct}. The framework consists of three components: a speech encoder, a quantizer, and a CTC decoder. The speech encoder converts the input speech signals into continuous representations, which are fed into the quantizer. The quantizer generates speech units and the corresponding codebook vectors. During the fine-tuning stage, the codebook vectors are passed to the CTC-based decoder. The parameters of all three components are jointly updated. During speech synthesis, a speech synthesizer is separately trained to convert discrete speech units from the quantizer into speech signals.

\subsection{Speech encoder}

We deploy SPIRAL \cite{huang2022spiral} as the SSL-based speech encoder.
%due to its efficiency and effectiveness. 
%SPIARL works by learning the perturbation-invariant representations of speech signals in a teacher-student framework. 
%By applying the in-utterance contrastive loss as a pre-training objective and imposing position randomization on the input to the teacher model, the content information of a speech utterance can be fully captured by the learned speech representation.
SPIRAL demonstrates competitive performance to wav2vec 2.0 \cite{baevski2020wav2vec}. While HuBERT \cite{hsu2021hubert} achieves impressive performance with the empirical determination of optimal embedding layers for different tasks, we prefer convenience and simplicity by selecting the output of the final layer of SPIRAL as direct input to the quantization module. We only require high-level speech content information for modeling tone-aware units, which SPIRAL is capable of. Furthermore, SPIRAL is configurable to encode latent speech presentations at different frame rates of 20, 40, and 80~ms, providing the flexibility for experimenting with the effect of different frame rates on speech discretization.

\subsection{Quantizer}

% [1] wav2vec2.0
% [2] vq-wav2vec
% [3] finite scalar quantization: vq-vae made simple
% [4] categorical reparameterization with gumbel-softmax
% [5] A * Sampling
% [6] statistical theory of extreme values and some practical applications: a series of lectures

\subsubsection{Gumbel-Softmax-based vector quantization (VQ)}
The VQ implementation is similar to that in \cite{baevski2020wav2vec}, except that only one codebook is used in our experimental setting. The Gumbel-Softmax enables choosing discrete codebook entries in a fully differentiable manner \cite{maddison2014sampling, jang2016categorical}. We also apply the straight-through estimator proposed in \cite{jang2016categorical}. The quantizer replaces the speech encoder output $\mathbf{z}$ by $\mathbf{\widehat{z}}=\mathbf{E}_i$ from a fixed-size codebook $\mathbf{E}\in \mathbb{R}^{V\times d}$ which contains $V$ vectors of size $d$. Specifically, the speech encoder output $\mathbf{z}$ is mapped to $\mathbf{l}\in \mathbb{R}^V$ logits, and the probabilities for choosing $j$-th codebook entry are computed by Equation~\ref{equation:eq1},

\begin{align}
  p_j=\frac{\exp(l_j+v_j)/\tau}{\sum_{k=1}^{V}\exp(l_k+v_k)/\tau}
  \label{equation:eq1}
\end{align}

where $\tau$ is a non-negative temperature, $v=-\log(-\log(u))$ and $u$ are uniform samples from $\mathcal{U}(0,1)$.

During the forward pass, the code index $i$ is chosen by $i=\arg\max_jp_j$, and in the backward pass, the true gradient of Gumbel-Softmax outputs is used.

\subsubsection{Finite scalar quantization (FSQ)}

% FSQ is proposed in \cite{mentzer2023finite} 
Mentzer et al. \cite{mentzer2023finite} propose FSQ as an alternative to VQ in the latent representation of VQ-VAE \cite{van2017neural}. In this approach, the VQ representation is reduced to very low dimensions. Each dimension is quantized into a limited set of discrete values, resulting in an implicit codebook formed by the product of these sets. Specifically, the speech encoder output $\mathbf{z}$ is first projected to a low-dimensional vector $\mathbf{\overline{z}}\in \mathbb{R}^n$, where $n$ is typically less than $10$. To quantize $\mathbf{\overline{z}}$ to a finite set of codewords, each entry $\mathbf{\overline{z}_m}$ is mapped to one of $L$ unique values by $\mathbf{\overline{z}_m} \mapsto \lfloor L/2 \rfloor \tanh (\mathbf{\overline{z}_m})$ followed by rounding to integers. Thereby, a quantized $\mathbf{\widehat{z}} \in \mathcal{C}$ is obtained. $\mathcal{C}$ represents the implied codebook, constituted by the product of these per-entry codebook sets, with $|\mathcal{C}|=L^n$. Taking $n=3$ and $L=3$ as an example, the codebook $\mathcal{C}$ is in the form $\{(-1,-1,-1),(-1,-1,0),(-1,-1,1),...,(1,1,1)\}$, where $|\mathcal{C}|=L^n=27$ \cite{mentzer2023finite}. The vectors in $|\mathcal{C}|$ can be enumerated, leading a bijection from any $\mathbf{\overline{z}}$ (i.e., speech encoder output $\mathbf{z}$) to an integer in $\{1,2,...,L^n\}$. Each entry of $\mathbf{\overline{z}}$ could be mapped to different $L_m$ values, thus the size of the codebook $\mathcal{C}$ is calculated as $|\mathcal{C}|=\prod_{m=1}^n L_m$.

% As stated in \cite{mentzer2023finite}, 
A straight-through estimator \cite{bengio2013estimating} is used to get gradients from the rounding operation to the encoder, similar to that in VQ-VAE. FSQ can use all codewords without relying on any auxiliary losses, subsequently avoiding the codebook collapse in VQ. Benefiting from very high codebook usage, FSQ can use large codebooks for better reconstruction quality.

\subsection{CTC fine-tuning}

The quantized representation $\mathbf{\widehat{z}}$ of speech encoder output $\mathbf{z}$ is fed into the CTC decoder. For tonal languages such as Mandarin Chinese, the CTC target is tonal phone sequence corresponding to input speech utterance, which provides tone information as supervision. For non-tonal languages, the training target is the original non-tonal phone sequence. During fine-tuning, the parameters of the speech encoder, quantizer (if applicable), and decoder are jointly updated. %The fine-tuning loss is defined as (1) the combination of CTC loss and codebook diversity loss used in \cite{baevski2020wav2vec} for VQ or (2) only CTC loss for FSQ.
For VQ, the fine-tuning objective is the combination of CTC loss and code-book diversity loss proposed in \cite{baevski2020wav2vec}. For FSQ, the fine-tuning objective is the CTC loss only.

\subsection{Speech synthesizer}

We choose VITS \cite{kim2021conditional} as the speech synthesizer in this study.
%This study uses VITS \cite{kim2021conditional} to transform discrete speech units into corresponding speech waveform. VITS %is a parallel text-to-speech system designed to perform both learning and synthesis in an end-to-end manner. It features a stochastic duration predictor to capture the varied rhythms in speech that text cannot represent. This innovation 
VITS directly synthesizes naturally sounded speech waveforms from extracted discrete speech units in an end-to-end manner without an external duration predictor. For a comprehensive understanding of this system, we refer readers to \cite{kim2021conditional}. 
To shorten the input sequence length of speech utterance, we further apply de-duplication \cite{lee2021direct,chang2023exploration} which merges consecutively repeated discrete speech units into a single unit.

\section{Experimental setup}
\label{sec:setup}
\subsection{Data}

% the duration of dev, test set of aishell-1 are different from paper
For Mandarin Chinese, we perform self-supervised learning on the SPIRAL speech encoder with 10000 hours of speech data from WenetSpeech \cite{zhang2022wenetspeech}. We utilize the 150-hour training set of AISHELL-1 \cite{bu2017aishell} to fine-tune the ToneUnit and train the VITS speech synthesizer. Since the speech synthesizer is trained on discrete speech units generated by the ToneUnit, using the same dataset for both systems is expected to eliminate data mismatch effects, thereby allowing our experiments to focus on obtaining tone-aware discrete speech units at this stage. The development and test sets of AISHELL-1 are used for checkpoint selection during fine-tuning and evaluation of synthesized speech, respectively. The CTC targets, which consist of 171 Mandarin phonemes with tonal markers, are prepared by transforming text with a Chinese grapheme-to-phoneme (G2P) package \cite{park2020g2pm}.

For English, we perform SSL on SPIRAL with 960 hours of speech data from LibriSpeech \cite{panayotov2015librispeech}. We apply the 100-hour subset (train-clean-100) for ToneUnit fine-tuning and VITS training with the same considerations as mentioned above. The checkpoint selection and synthesis evaluation are performed based on dev-other and test-clean, respectively. A total of 69 phonemes \cite{g2pE2019} are used as CTC targets.

% https://github.com/lingjzhu/charsiu

\subsection{Implementation details}
\label{subsec:imple}

The frame rate of the speech encoder is set to 20~ms. The Gumbel-Softmax quantizer generally follows the configuration used in wav2vec 2.0 \cite{baevski2020wav2vec}, but with a single codebook of 512 vector dimensions. The codebook size is set to 1000. For FSQ, the vector dimension is fixed at 4. The quantization level for each dimension is defined as \text{$\big[8, 5, 5, 5\big]$} to match the codebook size 1000. A linear layer is applied to obtain the low-dimensional projection from the speech encoder output. The CTC decoder module comprises 4 identical convolutional layers, each with a kernel size of 5, a stride size of 1, and ReLU as activation.

The AdamW optimizer \cite{loshchilov2017decoupled} is employed with a learning rate of \text{$3\times 10^{-5}$} for fine-tuning. The batch size is set to 8 for the Mandarin Chinese setting and 4 for the English setting. The fine-tuning process lasts for 320 epochs. The checkpoints with the lowest phone error rate on development sets are selected for extracting discrete speech units. VITS is trained for 100k updates following the training procedures and hyper-parameters in \cite{kim2021conditional}. Model evaluation is performed on the reconstructed audio based on the discrete speech units for the test set.

\subsection{Evaluation metrics}

% To measure the ability of discrete speech units to capture content information of speech signals in different scenarios, such as speaker, context, and environment, we propose to use Self-BLEU [ref] at the word level as the evaluation metric. For Chinese, we select $96$ two-character words with high occurrences in the AISHELL3 corpus [ref]. $100$ speech utterances for each word are sampled to make sure that each utterance occurs in different sentences or spoken by different speakers. The Self-BLEU among the $100$ utterances for each word is calculated. The scores for the $96$ words are averaged as the final result. A similar method is applied to construct a Self-BLEU test set for English. Specifically, $54$ words with a high occurrence rate are selected from the corpus of LibriSpeech-960 [ref]. $100$ different speech utterances are sampled. The average value of Self-BLEU of the $54$ words is reported. 

We track three metrics, including
\begin{itemize}
    \item Codebook Usage: The fraction of the codewords that are used at least 10 times during encoding the test set.
    \item Character Error Rate (CER) for Mandarin Chinese, or Word Error Rate (WER) for English: The metrics are used to assess the machine intelligibility of synthesized speech. For both languages, we use Whisper-large-v3 \cite{radford2023robust,whisperlargev3} to perform speech recognition on the synthesized audio of the test set.
    % \footnote{https://huggingface.co/openai/whisper-large-v3}
    % This metric evaluates the effectiveness of discrete speech units in capturing the content information of speech signals.
    \item Mean Opinion Score (MOS): The scores are collected from human listening tests to evaluate the naturalness of synthesized speech. 10 utterances are randomly sampled from the test set for each system. Each sample is rated by 20 raters on the scale: 1 (Bad), 2 (Poor), 3 (Fair), 4 (Good), 5 (Excellent).
\end{itemize}

\subsection{Baseline}

% \footnote{https://huggingface.co/TencentGameMate/chinese-hubert-base}
Applying k-means clustering to speech representations from SSL models such as HuBERT is a popular practice for generating discrete speech units \cite{lee2021direct, lakhotia2021generative,lee2021textless,polyak2021speech}. Consequently, we replicate this approach in our experiments using identical datasets as in Section \ref{subsec:imple}. Specifically, we utilize publicly available Mandarin Chinese HuBERT \textsc{Base} \cite{chinesehubertbase}, which is pre-trained on 10000-hour WenetSpeech Corpus. We re-train the k-means clusters with the AISHELL-1 training set, using the latent representations obtained from the \text{$9^{th}$} transformer layer of the model. For English speech, the publicly available HuBERT \textsc{Base} \cite{englishhubertbase}, which is pre-trained on the 960-hour LibriSpeech, is employed. We also re-train the k-means clusters with train-clean-100, also using the hidden embeddings from the \text{$9^{th}$} transformer layer. It should be noted that with k-means clustering, the centroids implicitly serve as a codebook. Thus, the number of k-means clusters is set to 1000 to match the above codebook size.
% \footnote{https://github.com/facebookresearch/fairseq/tree/main/examples/hubert}

To compare the effectiveness of k-means clustering in generating discrete speech units with the output of the last transformer layer of SPIRAL, we also include experimental results of SPIRAL with 1000-cluster k-means on the same datasets. 

Given the research's focus on obtaining discrete speech units capable of capturing tone information, a comparative analysis between discrete units from quantizer and continuous embeddings from speech encoder in speech synthesis is not conducted at this stage, as it falls outside the scope of this paper.

\section{Results and discussion}
\label{rec:result}
% The highest self-BLEU score, $84.4$, is recorded for the SPIRAL-VQ-500 model (5)
% Codebook sizes of $500$ and $1000$ are explored across various model settings, including speech encoders and quantizers, for the Chinese corpus. The evaluation results are detailed in Table~\ref{tab:zh_metric}. A larger codebook size results in a reduced self-BLEU score within identical model configurations comprising the same speech encoder and quantizer. When comparing models with equivalent codebook sizes, SPIRAL consistently has higher self-BLEU scores than HuBERT (1 vs. 3, 2 vs. 4). This indicates that unit sequences derived from SPIRAL's latent features demonstrate greater consistency across different speech contexts for the same content than those from HuBERT. Moreover, a comparison between VQ and FSQ (5 vs. 7, 6 vs. 8) reveals that reduced code utilization enhances self-BLEU scores. Consequently, employing fewer discrete units for speech signal representation appears to yield more stable tone-aware unit representations across different speaking conditions.

\subsection{Evaluation on speech synthesis}
% For Mandarin Chinese speech synthesis, we investigate codebook sizes of 500 and 1000 across different model configurations. %For 4-dimensional vectors in FSQ, the quantization level per dimension is defined as \text{$[4, 5, 5, 5]$} to match the codebook size 500 and as \text{$[8, 5, 5, 5]$} to match the codebook size 1000. 
% A larger codebook size consistently decreases CER within the same model setup, i.e., identical speech encoders and quantizers. When comparing VQ and FSQ models with equivalent codebook sizes, 
% It is observed that FSQ models exhibit better codebook utilization and achieve lower CERs than their VQ counterparts (ID 5 vs. 7, 6 vs. 8). Therefore, we suggest that increasing codebook usage potentially enhances intelligibility of speech synthesized from discrete units.
The evaluation results across different model configurations are summarized in Table~\ref{tab:zh_metric} for Mandarin Chinese. When we perform quantization with k-means, the CER of synthesized speech is much higher with substantially lower MOS results (ID 1 and 2). We notice that with 2 additional iterations of k-means clustering refinement in HuBERT, the setting with the HuBERT speech encoder outperforms that with the SPIRAL speech encoder. For the settings with the SPIRAL speech encoder, the CER decreases significantly with VQ or FSQ-based quantizers. The improvement of machine intelligibility of synthesized speech agrees with the improvement of corresponding MOS (ID 3 and 4). Notably, the FSQ setting (ID 4) exhibits a higher codebook utilization than VQ (ID 3) and achieves the best performance, with the lowest CER and the highest MOS.
%The MOS of synthesized speech using k-means as the quantizer (ID 2 and 4) is substantially lower than that of the original speech. 
%In contrast, both VQ and FSQ models (6 and 8) significantly improve the naturalness of the synthesized speech.
%The results suggest that intelligibility and naturalness of synthesized Mandarin Chinese speech are greatly affected by the "tone shift" issue when utilizing k-means for generating units; (2) both VQ and FSQ can effectively mitigate the "tone shift" problem. 

\begin{table}[thb]
\caption{Evaluation results for various model settings in Mandarin Chinese synthesized speech. The 95\% confidence interval for MOS is on average 0.14. GT corresponds to Groundtruth.}
\vspace{0.4cm}
\label{tab:zh_metric}
\centering
\resizebox{1.0\linewidth}{!}{
\begin{tabular}{c|cc|ccc}
\hline
ID & \begin{tabular}[c]{@{}c@{}}Speech\\encoder\end{tabular} & Quantizer & \begin{tabular}[c]{@{}c@{}}Codebook\\usage$\uparrow$ (\%)\end{tabular} & \begin{tabular}[c]{@{}c@{}}CER$\downarrow$\\(\%)\end{tabular} & MOS$\uparrow$       \\ \hline
   & GT                                                       & -         & -                                                             & 6.51                                               & 4.15\textpm0.55 \\ \hline
1  & HuBERT                                                   & k-means   & 100                                                           & 18.18                                              & 2.69\textpm0.55 \\
2  & SPIRAL                                                   & k-means   & 100                                                           & 25.76                                              & 1.99\textpm0.52 \\
3  & SPIRAL                                                   & VQ        & 74.4                                                          & 13.26                                              & 3.16\textpm0.62 \\
4  & SPIRAL                                                   & FSQ       & 100                                                           & \textbf{12.93}                                              & \textbf{3.55\textpm0.55} \\ \hline
\end{tabular}
}
\end{table}

\begin{table}[thb]
\caption{Evaluation results for various model settings in English speech. The 95\% confidence interval for MOS is on average 0.14. GT corresponds to Groundtruth.}
\vspace{0.4cm}
\label{tab:en_metric}
\centering
\resizebox{1.0\linewidth}{!}{
\begin{tabular}{c|ccccc}
\hline
ID & \begin{tabular}[c]{@{}c@{}}Speech\\encoder\end{tabular} & Quantizer & \begin{tabular}[c]{@{}c@{}}Codebook\\usage$\uparrow$ (\%)\end{tabular} & \begin{tabular}[c]{@{}c@{}}WER$\downarrow$\\ (\%)\end{tabular} & MOS$\uparrow$       \\ \hline
   & GT                                                       & -         & -                                                             & 1.84                                               & 4.52\textpm0.53 \\ \hline
5  & HuBERT                                                   & k-means   & 99.5                                                          & 5.03                                               & 3.68\textpm0.62 \\
6  & SPIRAL                                                   & k-means   & 99.9                                                          & 4.91                                               & 3.37\textpm0.49 \\
7  & SPIRAL                                                   & VQ        & 13.0                                                          & 7.46                                               & 3.20\textpm0.44 \\
8  & SPIRAL                                                   & FSQ       & 100                                                           & 4.31                                               & 3.64\textpm0.46 \\ \hline
\end{tabular}
}
\end{table}

Table~\ref{tab:en_metric} presents the evaluation results for English synthesis. The VQ model (ID 7) exhibits the worst results in terms of WER and MOS. We suspect that the VQ quantizer suffers from code collapse as the code utilization rate is only 13\%, which affects speech signal reconstruction adversely. The FSQ model (ID 8) achieves the lowest WER and attains MOS comparable to that of the HuBERT with k-means (ID 5). This indicates that FSQ is effective in addressing the codebook collapse issue and achieves more robust performance, which aligns with the findings reported in \cite{mentzer2023finite}. 

The above results indicate that, on the one hand, k-means clustering is capable of obtaining discrete speech units of non-tonal languages such as English speech for synthesizing intelligible and natural speech. On the other hand, a performance gap is observed when speech units from k-means clustering are used for synthesizing Mandarin Chinese speech. Our proposed ToneUnit framework improves speech recognition accuracy and MOS of synthesized Mandarin Chinese speech. The results suggest that leveraging text with tonal annotation for deriving discrete speech units is a remedy to the ``tone shift" issue. 

% It should be noted that within the proposed ToneUnit framework, the choice of speech encoder is flexible, extending beyond SPIRAL to potentially include HuBERT or other pre-trained speech models.

%The experimental results emphasize the importance of explicit tonal guidance in creating discrete speech units for tonal languages such as Mandarin Chinese. Without such supervision, k-means clustering tends to produce discrete units that exhibit the "tone shift" issue in synthesized Chinese speech. Our proposed ToneUnit framework successfully leverages text with tonal annotation to generate discrete units that overcome this issue. Consequently, the synthesized speech closely mirrors the original in terms of both intelligibility and naturalness. Moreover, FSQ outperforms VQ on these metrics while having a more simplified design.

\subsection{Representation of phonemes and tones via discrete speech units}

\begin{table}[thb]
\caption{Top-3 discrete speech Units for each tone of phonemes /a/, /i/, and /o/. The results are obtained on the AISHELL-1 test set using the FSQ model (ID 4).}
\vspace{0.4cm}
\label{tab:mapping}
\centering
\resizebox{0.9\linewidth}{!}{
\begin{tabular}{c|c|c}
\hline
Phoneme              & Tone & Top-3 speech units \\ \hline
\multirow{4}{*}{/a/} & 1    & 806; 815; 807      \\
                     & 2    & 4; 3; 9            \\
                     & 3    & 239; 439; 39       \\
                     & 4    & 838; 837; 829      \\ \hline
\multirow{4}{*}{/i/} & 1    & 600; 601; 800      \\
                     & 2    & 8; 0; 16           \\
                     & 3    & 38; 30; 29         \\
                     & 4    & 834; 826; 833      \\ \hline
\multirow{4}{*}{/o/} & 1    & 660; 651; 747      \\ 
                     & 2    & 270; 356; 262      \\
                     & 3    & 278                \\
                     & 4    & 876; 835; 875      \\
\hline
\end{tabular}
}
\end{table}

Since the codebook size, i.e., the number of available discrete speech units, is significantly larger than the number of phonemes, each phoneme is represented by a set of speech units. To investigate whether different tones of the same phoneme are represented by distinct sets of speech units, we map phonemes to speech units for each speech utterance in the AISHELL-1 test set using the FSQ model (ID 4). First, the phonemes output from the CTC decoder and the speech units output from the quantizer at the frame level for each utterance are collected. Subsequently, we count the corresponding speech units and their occurrences for each phoneme. 

We observe that each of the 171 phonemes is represented by a set of speech units, and conversely, one unit may correspond to several phonemes. For instance, the top-3 speech units with the highest occurrences for the phonemes /a/, /i/, and /o/ are shown in Table \ref{tab:mapping}. It is noteworthy that different tones of the same phoneme are represented by different sets of speech units. For example, Tone 1 and Tone 3 of the vowel /i/ can be distinguished by different speech unit representations despite their similarity in pitch distribution (as shown in Figure \ref{fig:tonalph}). A similar phenomenon is observed for Tone 2 and Tone 4 of /i/. This clearly demonstrates that tone differences are well captured by the discrete speech units generated by ToneUnit.

\subsection{Ablation study}

\begin{table}[thb]
\caption{Codebook usage and CER for different sizes of labeled data. The models employ SPIRAL as the speech encoder and FSQ as the quantizer.} %with a codebook size of 1000.
\vspace{0.4cm}
\label{tab:subset_metric}
\centering
\resizebox{0.9\linewidth}{!}{
\begin{tabular}{c|cc}
\hline
Labeled data & Codebook usage$\uparrow$ (\%) & CER$\downarrow$ (\%) \\ \hline
1 h    & 100                 & 15.00        \\
10 h   & 100                 & 13.03        \\
50 h   & 100                 & 13.04        \\ \hline
\end{tabular}
}
\end{table}

A con of ToneUnit is the requirement of annotated data with tone labels. To evaluate the required amount of labeled data, we perform ablation under the same configuration as the FSQ model (ID 4) with different sizes of the AISHELL-1 training set. We randomly sample 3 training subsets of 1, 10, and 50 hours from the full 150-hour training set for the fine-tuning process. We still train the speech synthesizer with the full 150-hour AISHELL-1 dataset for 100k updates. Note that the speech synthesizer is trained with extracted discrete speech units without any text labels. The results are reported in Table \ref{tab:subset_metric} in terms of codebook usage and CER. FSQ can still fully use the codewords at different sizes of labeled data. With just 1 hour of annotated data with tone labels, the ToneUnit framework successfully generates speech units capable of synthesizing Mandarin Chinese speech at acceptable machine intelligibility. With a subset of 10-hour training data, the CER of synthesized speech is already comparable to those from the full training set. This suggests that the proposed ToneUnit framework retains effectiveness even with limited labeled data. 

%The results suggest that intelligibility and naturalness of synthesized Mandarin Chinese speech are greatly affected by the "tone shift" issue when utilizing k-means for generating units; (2) both VQ and FSQ can effectively mitigate the "tone shift" problem. 

\section{Conclusion}
\label{sec:Conclusion}
This study proposes the ToneUnit framework, which addresses the ``tone shift"  issue in synthesized Mandarin Chinese speech based on discrete speech units. By leveraging text data annotated with tone labels as CTC supervision, ToneUnit is able to produce discrete speech units with better phonemic tone discrimination in Mandarin Chinese. Experimental results demonstrate that speech units generated by ToneUnit can synthesize intelligible and natural speech for both Mandarin Chinese and English. Additionally, FSQ is a more effective quantization method to produce tone-aware speech units than VQ, with a much simpler structure. Moreover, ToneUnit works with a limited amount of annotated data. We hypothesize that ToneUnit can serve as a connector between the speech modality and the large language model (LLM) backbone in multimodal LLMs. This potential application will be investigated in our future work.
% We presume that ToneUnit can be used as a general speech discretization method to learn context-aware discrete speech representations. 

% \pagebreak
\bibliographystyle{IEEEbib}
\bibliography{refs}

\end{document}